\documentclass[sigconf]{acmart}

\AtBeginDocument{%
  \providecommand\BibTeX{{%
    \normalfont B\kern-0.5em{\scshape i\kern-0.25em b}\kern-0.8em\TeX}}}

\copyrightyear{2024}
\acmYear{2024}
\setcopyright{acmlicensed}\acmConference[CHI '24]{Proceedings of the CHI Conference on Human Factors in Computing Systems}{May 11--16, 2024}{Honolulu, HI, USA}
\acmBooktitle{Proceedings of the CHI Conference on Human Factors in Computing Systems (CHI '24), May 11--16, 2024, Honolulu, HI, USA}
\acmDOI{10.1145/3613904.3642215}
\acmISBN{979-8-4007-0330-0/24/05}

\begin{document}

\title[PD-Insighter: Visualizing Daily Actions for Parkinson's Disease Treatment]{PD-Insighter: A Visual Analytics System to Monitor Daily Actions for Parkinson's Disease Treatment}

\author{Jade Kandel}
\email{kandelj@cs.unc.edu}
\affiliation{%
  \institution{University of North Carolina at Chapel Hill}
  \city{Chapel Hill}
  \state{North Carolina}
  \country{USA}
}

\author{Chelsea Duppen}

\email{chelsea_parker@med.unc.edu}
\affiliation{%
  \institution{University of North Carolina at Chapel Hill}
  \city{Chapel Hill}
  \state{North Carolina}
  \country{USA}
}

\author{Qian Zhang}
\email{qzane@cs.unc.edu}
\affiliation{%
  \institution{University of North Carolina at Chapel Hill}
  \city{Chapel Hill}
  \state{North Carolina}
  \country{USA}
}

\author{Howard Jiang}
\email{hhjiang2023@gmail.com}
\affiliation{%
  \institution{University of North Carolina at Chapel Hill}
  \city{Chapel Hill}
  \state{North Carolina}
  \country{USA}
}

\author{Angelos Angelopoulos}
\email{aangelos@cs.unc.edu}
\affiliation{%
  \institution{University of North Carolina at Chapel Hill}
  \city{Chapel Hill}
  \state{North Carolina}
  \country{USA}
}

\author{Ashley Neall}
\email{aneall@unc.edu}
\affiliation{%
  \institution{University of North Carolina at Chapel Hill}
  \city{Chapel Hill}
  \state{North Carolina}
  \country{USA}
}

\author{Pranav Wagh}
\email{pawagh@unc.edu}
\affiliation{%
  \institution{University of North Carolina at Chapel Hill}
  \city{Chapel Hill}
  \state{North Carolina}
  \country{USA}
}

\author{Daniel Szafir}
\email{dszafir@cs.unc.edu}
\affiliation{%
  \institution{University of North Carolina at Chapel Hill}
  \city{Chapel Hill}
  \state{North Carolina}
  \country{USA}
}

\author{Henry Fuchs}
\email{fuchs@cs.unc.edu}
\affiliation{%
  \institution{University of North Carolina at Chapel Hill}
  \city{Chapel Hill}
  \state{North Carolina}
  \country{USA}
}

\author{Michael Lewek}
\email{michael_lewek@med.unc.edu}
\affiliation{%
  \institution{University of North Carolina at Chapel Hill}
  \city{Chapel Hill}
  \state{North Carolina}
  \country{USA}
}

\author{Danielle Albers Szafir}
\email{danielle.szafir@cs.unc.edu}
\affiliation{%
  \institution{University of North Carolina at Chapel Hill}
  \city{Chapel Hill}
  \state{North Carolina}
  \country{USA}
}

\renewcommand{\shortauthors}{Kandel, et al.}

\begin{abstract}

\begin{figure*}[h]
  \centering                   \includegraphics[width=\textwidth]{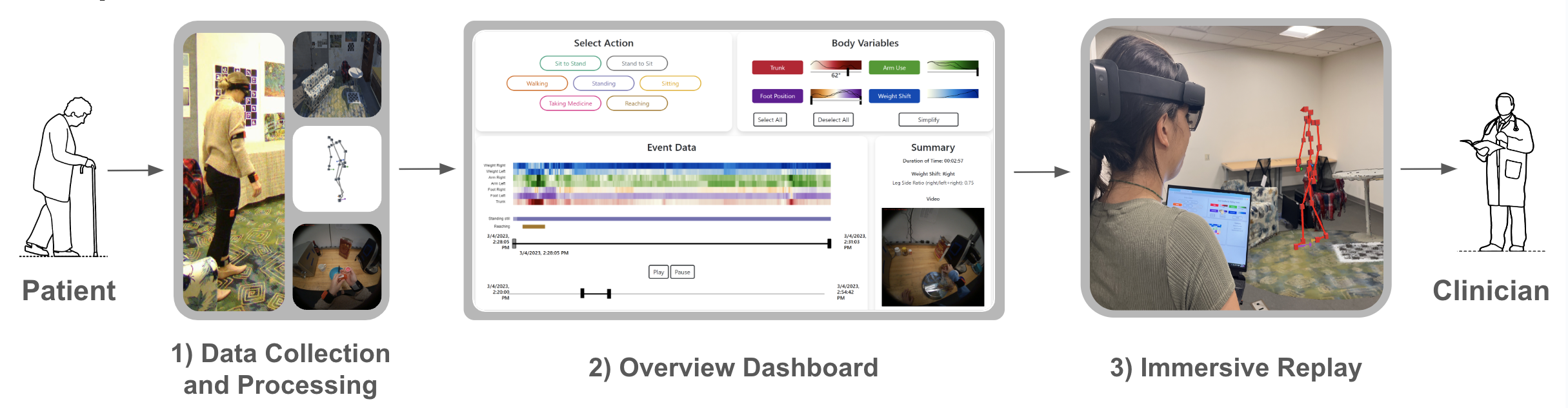}
    \caption{We present PD-Insighter, a visual analytics system for assisting clinicians treating Parkinson's Disease. Our system presents (1) relevant patient body motion, environment, and behavioral information extracted from  video and sensor data in both (2) an overview dashboard and (3) an immersive replay that enables clinician interaction and analysis of patient data.
    }

    \Description{Figure 1 illustrates the PD-Insighter system workflow through a streamlined flow diagram. Initially, it features an image of a patient, marking the start of the process. This progresses to the second stage, "Data Collection and Processing," where patient-related information is gathered and processed. Following this, the "Overview Dashboard" is presented, showcasing the processed data in an accessible format. The next step introduces the "Immersive Replay," depicted by an individual wearing an augmented reality headset. This person observes a digitally-rendered skeleton within a virtual environment, simultaneously managing the Overview Dashboard on a laptop. The workflow concludes with a clinician receiving the visualization, symbolized by an image of a healthcare professional. }

  \label{fig:teaser}
\end{figure*}

People with Parkinson's Disease (PD) can slow the progression of their symptoms with physical therapy. However, clinicians lack insight into patients' motor function during daily life, preventing them from tailoring treatment protocols to patient needs. This paper introduces \textit{PD-Insighter}, a system for comprehensive analysis of a person's daily movements for clinical review and decision-making. PD-Insighter provides an overview dashboard for discovering motor patterns and identifying critical deficits during activities of daily living and an immersive replay for closely studying the patient's body movements with environmental context. Developed using an iterative design study methodology in consultation with clinicians, we found that PD-Insighter's ability to aggregate and display data with respect to time, actions, and local environment enabled clinicians to assess a person's overall functioning during daily life outside the clinic. PD-Insighter's design offers future guidance for generalized multiperspective body motion analytics, which may significantly improve clinical decision-making and slow the functional decline of PD and other medical conditions.

\end{abstract}

\ccsdesc[500]{Human-centered computing~Visual analytics}

\keywords{Visualization, Health, Immersive Analytics}

\maketitle

\section{Introduction}

Individuals with Parkinson's Disease (PD) develop various motor deficits, including gait disturbances and balance impairments, as the disease progresses \cite{moustafa_motor_2016}. Although there is no cure for PD, physical therapy can potentially increase movement capacity and create long-term improvements in many motor impairments and functional activities \cite{grabli_normal_2012}. Tailoring therapy protocols requires insight into how individuals perform various motor activities over time. Clinicians currently only have access to information about a patient's movements through brief, periodic clinical sessions and subjective self-reporting. Unfortunately, people with PD often struggle to remember moments of movement difficulty, 
and their written diaries can be unreliable due to recall bias and non-compliance \cite{stone_patient_2002}, necessitating quantifiable measures of mobility to adequately tune physical therapy protocols.



We introduce \textit{PD-Insighter}, a system for investigating in-home body motion data to support clinician understanding of patient motion patterns over time. PD-Insighter leverages motion capture data, activity labels, and environment videos to provide a multifaceted view of body movement, enabling interactive visualization for several hours of body motion capture. 
%
We designed PD-Insighter to satisfy three primary goals elicited through interviews with our 
clinical collaborators: 1) compute key indicators of body movement reflective of stability, balance, and gait; 2) aggregate and summarize in-home activity data to support rapid identification of relevant time segments and patterns for review; and 3) provide methods for closely studying instances, events, and activities of interest. PD-Insighter allows analysts to explore patient behavior simultaneously across \textit{body variables}, consisting of numeric body displacements and angles of interest, and \textit{action labels}, consisting of categorical labels indicating the action performed at a certain time, such as walking, sitting, or taking medicine.


PD-Insighter uses a hybrid-platform design, with a desktop-based \textit{Overview Dashboard} to present broad patterns and an \textit{Immersive Replay} showing body and environment reconstructions in AR 
for contextualized, detailed analysis. 
The Overview Dashboard enables action-driven analysis 
using temporal heatmaps designed to emphasize outliers, display trends, and summarize body motion data. 
Immersive Replays, which reconstruct patient motion and physical environment data from raw data captures, provide clinicians the ability to analyze timepoints reflecting potential motor challenges in greater detail. Given that clinicians are trained to study the patient’s body in-situ, we leveraged augmented reality's spatial rendering, navigation, and life-size representation to mirror current in-person clinical practices.
By coupling immersive and traditional visual analytics, we enable close investigation with AR while preserving intuitive interactions with the dashboard for high-level navigation. 

We evaluated our system in a think-aloud study with six rehabilitation specialists. We found that PD-Insighter enabled rapid and effective insight into motion data across multiple levels of detail. This system provides preliminary steps towards a broad vision for enhanced therapies for individuals with PD. Enhanced feedback for physical, observable tasks may optimize treatment protocols and combat the progression of PD as well as 
other chronic diseases such as stroke.

\textbf{Contributions:} Our primary contribution is a visual analytics approach for analyzing motion data in PD treatment embodied in \emph{PD-Insighter}, a system for exploring
body motion data. In creating the system, we contribute the following:
\begin{itemize}
  \item a task characterization of clinical motion analysis tasks in physical therapy applications
  \item a data processing pipeline for synthesizing key measures for PD motion analysis from traditional motion capture data
  \item a hybrid visual analytics design that combines traditional and immersive analytic methods for simultaneous body motion, action, and video data analysis.
\end{itemize}


\section{Related Work}
PD-Insighter leverages visual analytics to support clinicians in supporting patients with PD. Our approach combines ideas from visual analytics in biomedical applications to analyze high-level motion patterns and from immersive analytics to provide contextualized, detailed replays of patient motion. 

\subsection{Visualizations for Biomedical Analysis}

Visual analytics tools have supported various clinical decision-making tasks, such as analyzing medical events over time \cite{zhang_idmvis_2019,plaisant_lifelines_2003}, determining demographic risk factors and related outcomes \cite{zhang_iterative_2015, rogers_composervisual_2019}, and monitoring longitudinal disease progression \cite{wang_threadstates_2022,weber_international_2021}. Using such tools in conjunction with remote patient monitoring systems, health experts can observe patients away from the hospital using various health metrics, including blood pressure, pulse, heart rate, body temperature, and oxygen levels \cite{malche_artificial_2022,ramesh_mobile_2012,dabbakuti_design_2012}. 
Remote monitoring approaches would significantly benefit people with PD, whose motor impairments can make clinical visits especially challenging \cite{dorsey2016moving}. Body-worn sensors can collect data about people's movement patterns for remote monitoring applications; however, data from such sensors tends to be insufficient for clinical analysis when used in isolation (see Del Din et al. \cite{del2021body} for a survey). 

Activities of daily living (ADLs) 
can provide context for interpreting body-worn sensor data. ADLs describe the essential abilities needed to self-sufficiently manage one's care, including eating, showering, and walking. Medical specialists consider ADLs crucial in diagnosing and treating conditions such as PD  \cite{alizadeh_telemonitoring_2011, deal_parkinsons_2019, mahoney_functional_1965}. People with PD may exhibit motor deficits differently across ADLs. For example, a pronounced forward lean is normal while reaching for an object in front of you but may indicate a postural impairment when walking. 
 
Understanding activity data during ADLs often requires a knowledge of the patient's environment to 
discover how objects 
in the home might play a role in movement patterns. The presence of a couch and dining table correlate with sitting and low activity while transitions between spaces correlate with walking and high activity \cite{zhongna_zhou_activity_2008}. Past studies of ADLs found that medical specialists benefit from contextualizing behavioral data based on the patient's location (e.g., kitchen, bedroom, bathroom) \cite{robben_identifying_nodate}. Other systems visualize ADLs with 
graphical representations of the patient's apartment, using trajectories or avatar replays to show movement in space  \cite{rashidi_mining_2010, boers_smart_2009}; however, these approaches only provide short snapshots  
with limited global insight into movements over time, requiring analysts to remember key patterns in ADL data throughout the replay. Gil et al. \cite{gil_data_2007} summarized activity within the home using embedded sensors on common objects to infer the kinds of activities people engaged in over time. While this approach provides knowledge about the number of actions occurring in various 
locations, it provides limited insight into movement quality. 

Other approaches enable clinicians to analyze specific biomechanical motions from a given activity in detail. 
Ploderer et al. \cite{ploderer_how_2016} visualized upper limb movement in people who had a stroke
using multiple representations of arm movement across time, the amount of time active, and heatmaps on 2D avatars showing body areas with high activity. 
HAExplorer \cite{eulzer_haexplorer_2022} 
displays precise helical components of biomechanical motion at individual joints. Keefe et al. \cite{keefe_interactive_2009} provided a small multiples view of biomechanical motions from a specific activity to view from multiple perspectives. However, these approaches focus on specific motions and joints over discrete activities in isolation rather than global, multi-activity perspectives and environmental contexts associated with motor abilities in daily life. Other systems used Kinect for data capture to create a full-body skeletal reconstruction and extract kinematic features, capturing 
in home body motion data for clinical analysis and PD diagnosis \cite{nguyen_mining_2017, kao_validation_2016}. However, for ADLs, skeleton reconstructions and kinematic values 
can be difficult to analyse without environment, temporal, or action context.

Motion analysis for PD and similar motor disorders must balance a crucial tension between access to high-level summaries showing how body motions change 
over time (i.e., hours to days) and low-level details enabling deeply contextualized analysis over brief movements or activities (i.e., seconds to minutes). Systems representing ADLs with activity level or location can present data across long time periods but lack environmental and full body contexts to draw meaningful conclusions about behavioral patterns. Replays of patient motions can provide essential context but only work for brief time segments. We pair statistical, activity-based overviews with deeply contextualized immersive reconstructions to enable an understanding of large body motion patterns across time and the analysis of context.

\subsection{Immersive Analytics for Body and Environment Data}

To help clinicians analyze patient activities in context, PD-Insighter combines traditional visualization workflows with immersive analytics (IA). IA leverages immersive technologies such as virtual and augmented reality (VR and AR) for data visualization \cite{marriott_immersive_2018}. Immersive environments project data into 3D space, often making navigating, manipulating, and perceiving spatial information such as depth, height, and size more intuitive in part by integrating binocular cues and by allowing people to physically move through data \cite{marriott_immersive_2018,whitlock_graphical_nodate, kraus_value_2021, zacks_reading_1998, heinrich2021estimating}. Improved spatial perception from AR is especially important to our work, as analysing 3D body limbs in space requires accurate understanding of depth and distance. 
Our work is particularly informed by research in situated analytics (SA), which directly integrates visual representations of the data into the studied environment, often placing data on top of physical objects such as buildings or bodies \cite{willett_embedded_2017,bressa_whats_2022}. SA can incorporate data directly into the physical environment, advancing fields such as fieldwork \cite{whitlock_designing_2019}, construction \cite{irizarry_infospot_2013}, and interior design \cite{wang_pointshopar_2023}. 

Past work in SA summarizes human motion within an indoor environment, preserving the time, place, and interactions with the space \cite{luo_pearl_2023,kloiber_immersive_2020}. For example, AvatAR \cite{reipschlager_avatar_2022} 
paired a digital avatar with different embedded representations of motion (e.g., gaze, paths) to reflect movement patterns and interaction with the environment. However, these systems 
primarily visualize
where people have interacted with the space rather than the quality of their movements. 
We instead focus our approach on understanding the mechanics of patient motion, where context and location play a vital secondary role in interpreting movement patterns. 

SA systems have also situated visualizations with respect to the body, supporting more detailed mechanical motion analyses, such as improving cycling performance \cite{kaplan_towards_2018}, basketball free throws \cite{lin_towards_2021}, hand gestures \cite{freeman_shadowguides_2009}, and physical rehabilitation \cite{doyle_base_2010,garcia_mobile_2014,ayoade_novel_2014}. Real-time systems such as Physio@Home \cite{tang_physiohome_2015} used AR mirrors to embed wedges and lines directly on the person's body to guide movement for upper limb physiotherapy. YouMove \cite{anderson_youmove_2013} also employed an AR mirror to teach physical movement sequences by having people follow along with a skeleton's actions. These systems demonstrate the power of IA for real-time motion analysis and guidance; however, they focus on short time-frames for guiding motion in real-time. 

While traditional visual analytics approaches can support coarse-grained longitudinal analysis, past work in IA suggests its potential for contextualized investigations and replays of body motion and key actions. Clinicians can use IA tools to understand the full context of a movement and view body motion from multiple angles to gain a holistic picture of potential motor deficits. Further, IA can offer visceral experiences with data \cite{lee_data_2021,zhou_data-driven_2023}, which may augment data interpretation by emphasizing relative proportions of the patient's physical space and associated objects and actions. We pair this paradigm with traditional overview analytics in PD-Insighter to enable multi-scale insight into movement data for physical therapy for PD.

\section{Task Analysis}
\label{sec:tasks}
We developed PD-Insighter using an approach drawing on the design study methodology \cite{sedlmair_design_2012}.
Our research team included two experts in neurologic physical therapy specializing in PD 
who helped identify key tasks 
for using data to enhance clinical understanding for therapy guidance. These experts helped refine the tasks and visualization approach during system development. 

\subsection{Motivating Problem \& Terminology}
People with PD develop various motor deficits, including gait disturbances and balance impairments that can lead to falls, gait freezes, 
and difficulties standing and 
grabbing an object \cite{moustafa_motor_2016}. To determine effective rehabilitative treatment methods, clinicians must efficiently and effectively identify motor deficits and their potential triggers. Triggers can include changes in the environment \cite{almeida_freezing_2010}, time since medication \cite{amboni_prevalence_2015}, and divided attention \cite{chomiak_new_2015}. 

Data for body motion during ADLs could provide clinicians with crucial information 
about deficits and triggers before meeting a patient. For example, if a person with PD expresses 
challenges standing up from their chair, a clinician could examine their data prior to a clinical visit to determine potential triggers causing this difficulty. The clinician could then spend the clinical visit educating the patient on exercises or compensatory methods to improve sit-to-stand transfers. Due to time constraints in the clinic, clinicians 
require rapid and efficient methods for reviewing this data.

To understand patient motor behavior, clinicians need to 1) focus on the patient's ability to perform specific ADLs (e.g. walking, standing, reaching), 2) discover when and how frequently motor deficits occur, and 3) understand movement context to see why motor deficits occur. While clinicians currently rely on brief in-clinic assessments and self-reports, data-driven approaches can provide more precise and holistic insight into patient movement. 

While the terminology may vary across practices, clinicians identified three kinds of data that would support clinical analysis and decision-making for PD:
\textit{body}, \textit{action}, and \textit{event}.
For this paper, \emph{body variables} are specific joint angles and displacements where changes in value can indicate a patient's motor performance. 
An \textit{action} is the process of moving one's body to achieve a goal, such as walking, sitting, or reaching. Actions can make up ADLs (e.g., reaching is performed while bathing). An \textit{event} is the time period an action occurs.

Clinicians can infer motor deficits from shifts in body variables; however, clinicians can only fully make sense of body variable changes in the context of performed actions. For example, a patient may stop moving due to a freeze or due to external stimuli such as searching the refrigerator. Thus, clinicians desired a way to contextualize body variables with respect to action data, temporal events, and the physical environment to readily enable clinicians to investigate motor deficits and their potential triggers.

\subsection{Tasks}
\label{tasks}
While clinicians do not currently have tools for working with longitudinal body motion and action data in PD treatment, we identified core tasks associated with such data through multiple clinician interviews. These tasks focused on data needs for understanding proper and improper body motion and what motion data indicates about the patient's condition. We revised these tasks through continued iteration with clinical collaborators on early system prototypes. 

We characterize these tasks into three high-level goals: \textit{identify and filter data by action}, where clinicians can assess movement patterns over key ADLs; \textit{discover motor deficits}, where clinicians can compare motion patterns across key body relationships to discover systemic deficiencies; and \textit{contextualize motor deficits}, where clinicians can view potential deficits in context to understand potential triggers and treatments. 

\subsubsection{Task 1: Identify and Filter Data by Action} 
\label{tasks_actions}

Clinicians focus on specific actions to understand a patient's challenges. Actions consist of complex combinations of movements that together enable an intended goal. For example, the ``walking'' action combines synchronized and balanced footsteps, arm swings, and upright posture to move from one place to another. The patterns clinicians look for in motion data depend on the patient's action. For example, significant variations in trunk angle during walking may indicate a deficit, whereas the same variations are normal when standing up from a chair. 

\begin{figure*}
  \includegraphics[width=\textwidth]{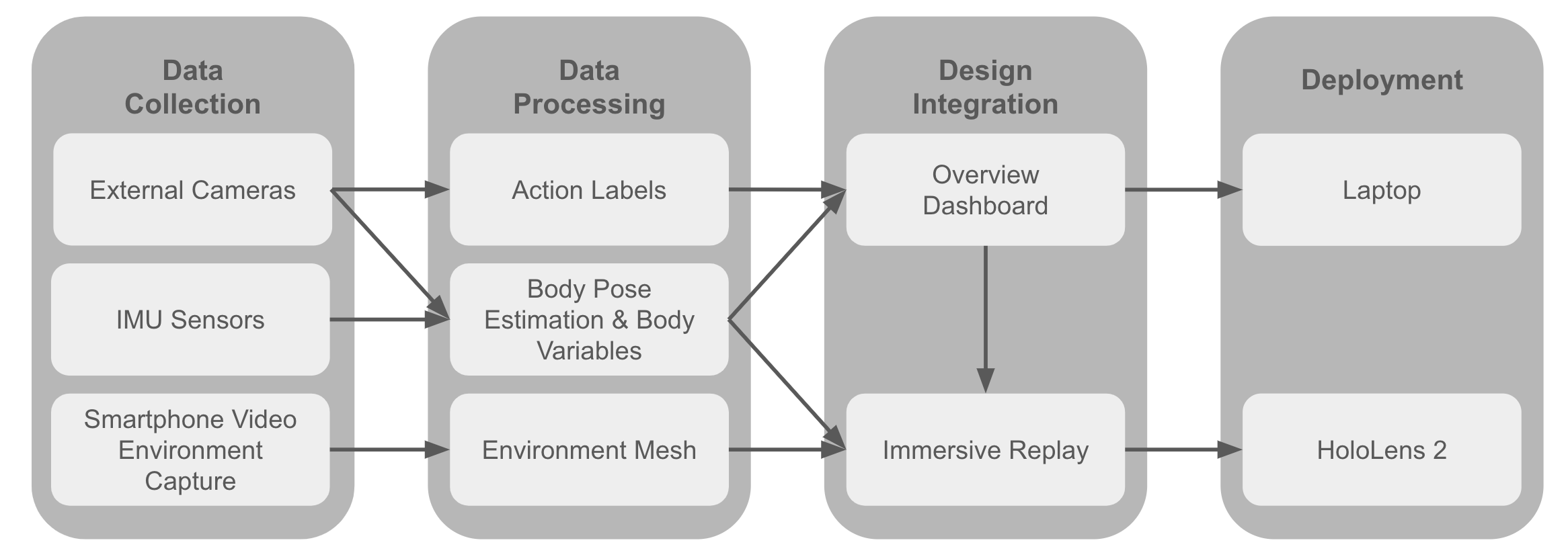}
  \caption{\textbf{Workflow Diagram}: 
  The pipeline and tools used to implement PD-Insighter. First, data collected about the 
  person and their environment is processed to extract body pose, action labels, and an environment mesh. 
  Next, PD-Insighter 
  presents this data to clinicians through the Overview Dashboard, which visualizes the body variables and action labels together, and the Immersive Replay, which visualizes the 
  body pose 
  and environment mesh together. 
}
    \Description{Figure 2 presents a detailed flowchart that outlines the sequential workflow of our system, transitioning from left to right across four key stages. 1) "Data Collection", which includes "External Cameras", "IMU Sensors", and "Smartphone Video". 2) "Data Processing", which includes "Action Labels", "Body Pose Estimation", and "Environment Mesh", and are derived from the collected data in the first stage. 3) "Design Integration" , which includes "Overview Dashboard" and "Immersive Replay", both designed to visualize and interact with the processed data from stage 2. 4)  "Deployment" with the dashboard on a "Laptop" and the replay on "HoloLens 2", illustrating the system's data journey from collection to application.}
  
    \label{fig:workflow}
\end{figure*}

Key actions for informing treatment protocols include:
\begin{itemize} 
\item Sit-to-stand (e.g., when a patient gets up from a chair)
\item Sitting 
\item Stand-to-sit (e.g., when a patient sits down on a chair)
\item Reaching (e.g., when a patient grabs something from a high shelf or across a table)
\item Walking
\item Standing
\item Taking medicine
\end{itemize} 

While other actions may also inform protocols, we focus on these common actions as our clinician collaborators identified these as necessary to understand a patient's condition. By letting clinicians examine motion patterns with respect to actions over time, clinicians can more readily identify and select time segments, filter and compare events of the same action, and select a specific event based on actions of interest (see \S \ref{overview}).
 
\subsubsection{Task 2: Discover Motor Deficits}
\label{sec:deficit_tasks}
Different body angles and displacements illuminate if, when, and why a motor deficit occurs. Clinicians frequently need to assess differences in posture, arm use, gait patterns, and weight balance (see \S \ref{body_variables} \& \ref{body-sliders}). 

\textbf{Posture:} Poor posture is a common symptom of PD and can lead to neck or back pain, improper breath control, loss of flexibility/mobility, and increased risk for falls from poor balance. The trunk angle---the angle between the vertical axis and vector between the pelvis and neck---captures posture, showing how far a patient's upper body is leaning.

\textbf{Arm Use:} Asymmetric arm swing during gait is a characteristic early motor sign of PD \cite{lewek_arm_2010}. As the disease progresses, arm swing is further reduced bilaterally. Understanding the context of arm swing is critical. An asymmetric arm swing may be a sign of PD, but it can also occur when someone is holding a glass of water and walking across the room. We can infer when one arm is swinging more or used more than the other by looking at patterns in the distance between each hand and the pelvis. 

\textbf{Gait Patterns:}
Clinicians examine patterns in foot position relative to the pelvis to 
estimate step lengths in gait. Changes in terrain or changes in mobility might cause variations in foot position patterns. Gait freezing is an abnormal walking pattern where patients are temporarily unable to move their feet forward when trying to walk, which can lead to a fall. A clinician can infer a  freeze event when the feet suddenly stop or strides grow unexpectedly short under the pelvis while walking. 

\textbf{Weight Balance:} Imbalanced weight shift can create postural instability and freezing \cite{dijkstra_impaired_2021} as well as struggles sitting down, standing up, or standing still. We can infer when one leg bears more weight based on the displacement between the pelvis and the foot from the side direction.  


With potentially hours' and days' worth of body motion data, finding motor deficits efficiently is challenging. To enable clinicians to find these important changes in motion quickly, we can aggregate data over time with respect to using meaningful statistics and actions and emphasize outliers that reflect a motor deficit's shift in body motion (see \S \ref{overview}).

\subsubsection{Task 3: Contextualize Motor Deficits}
\label{contextualize_deficits}
Time of day, symptom and action duration, and where the person was in their house can indicate what may have contributed to the observed symptoms. Clinicians need temporal context to assess the causes of motion change and identify potential triggers (see \S \ref{overview}). For example, if and when someone took their medicine may impact observed symptoms and motor patterns \cite{amboni_prevalence_2015}. 
How long someone froze or fell can indicate the severity of the event. The percentage of time sitting can show how much someone was active throughout the day, which may increase or decrease the likelihood of symptoms. 

Although joint angles and displacements can 
highlight potential points of concern, clinicians also need to study a patient's physical body and environment rather than rely solely on numerical representations. Direct observation provides contextual understanding, validates findings, and captures nuances that numbers may overlook, such as environmental obstacles. For instance, carpet/floor changes, doorways, and turns can trigger a freeze event. Patients might also rely on their surrounding environment for physical support, such as leaning on a table while walking or pressing off the couch when standing up (see \S \ref{immersive}).


\section{System Overview}

PD-Insighter is a hybrid desktop-AR system presenting body motion, action, and environment data for clinicians to investigate the motor functioning of a person with PD. 
Our system supports the target tasks elicited in our 
interviews through a workflow consisting of data collection, data processing, data integration, and deployment (Figure \ref{fig:workflow}).
We tested our system with datasets of 
various lengths and found our visualization methods supported up to five hours of analysis on a traditional 
laptop display. 
        
PD-Insighter takes body pose data, action labels, and video data as input. PD-Insighter processes that input to compute a series of body variables reflecting key joint and displacement relationships (see \S \ref{body_variables}) and reconstructs the local environment from video data (see \S \ref{data_environment_reconstruction}). PD-Insighter passes the processed data into the analytics component for clinical analysis.
    
The visual analytics components of PD-Insighter consist of a desktop-based Overview Dashboard (see \S \ref{overview}) presenting the action labels and body variables (Figure \ref{fig:overview_dashboard}) and an AR Immersive Replay (see \S \ref{immersive}) that provides detailed context for specific timepoints of interest (Figure \ref{fig:immersive_replay}). The dashboard first presents clinicians with a summary overview of actions and variable distributions over the entire dataset (Figure \ref{fig:data_view}.1). Clinicians can use action types, body variables, and times to drill down into the data, characterize broader movement patterns, and isolate potential timepoints of interest (Figure \ref{fig:data_view}.2 \& \ref{fig:data_view}.3). The clinician can select a moment in time to view the context of the movements in the Immersive Replay. The immersive display renders a skeleton from the body pose data situated in the environment reconstruction. We display the reconstruction of the local environment centered on the patient with a 1.5-meter radius view of the environment to provide context for understanding motor behavior and enable the clinician to view the reconstructed environment and patient movement from multiple perspectives. Clinicians can move back and forth between the Overview and Immersive Replay on demand.

\subsection{Implementation}
We implemented the dashboard using D3.js and HTML5. We built the Immersive Replay in Unity and deployed it to a HoloLens 2 headset. We managed data processing and reconstruction using custom Python scripts, EasyMocap \cite{noauthor_zju3dveasymocap_2023}, and RealityCapture\cite{noauthor_realitycapture_nodate}. 
\section{Data Collection and Processing}
 PD-Insighter requires body motion data consisting of body variables, action labels, and an environment mesh, which can be extracted from motion capture and standard video. Recent work in PD has deployed IMUs and similar body-worn sensors in the home for remote monitoring applications \cite{del2021body}. 
 Advances in egocentric motion capture and activity recognition lead clinicians to anticipate that full non-intrusive motion capture during ADLs using other wearable devices, such as cameras, 
 will soon be feasible \cite{zhang_reconstruction_2023, cha_mobile_2021,lin_learning_2022, wang_attentive_2019, demrozi_comprehensive_2023, kolkar_human_2023}. 
 Our current system 
 presents a proof-of-concept of such a pipeline, using external cameras and body-worn inertial measurement units (IMUs) similar to those employed by our clinical collaborators for in-lab observation. Here, we describe our current capture workflow; however, PD-Insighter can also use data generated through alternative pipelines, including conventional motion capture. For generalizability, the system uses conventional motion capture data formats (i.e., a JSON file summarizing joint positions for each frame) annotated with simple action labels (i.e., a JSON file with start and end frame numbers) and environment meshes generated through standard RGB video files.

\begin{figure}
  \includegraphics[width=\columnwidth]{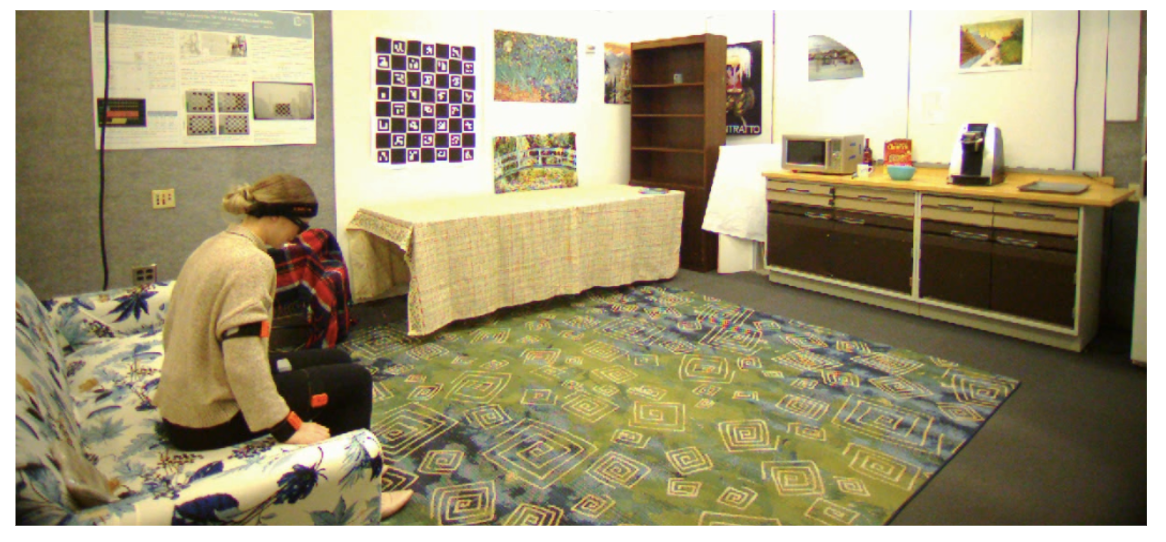}
  \caption{\textbf{Data Collection:} A person wearing IMUs acting out ADLs in our artificial apartment environment for data collection. 
  }
  \Description{A person with sensors strapped onto their arms and legs sitting on a couch. The person and couch are in an artificial apartment environment which includes a table, kitchen area, and shelf. }

    \label{fig:data_collection}
\end{figure}

\subsection{Data Collection}
To develop PD-Insighter, we collected body motion data from a combination of RGB cameras and IMUs, following practices used in current clinical observations conducted in Parkinson's research and similar studies \cite{burpee_biomechanical_2015}. 
We conducted our data collection
in a 11ft $\times$ 15ft environment, shown in Figure \ref{fig:data_collection}, which simulated a domestic living space (i.e., living room and kitchen) outfitted with everyday furniture items such as couches, counters, tables, chairs, and shelves and essential appliances including microwaves, coffee makers, utensils, and dishes. We captured the local environment using a smartphone video recording. The tracked person wore ten IMU sensors on their body to collect body pose data. Three large cameras on tripods record the person in the room performing actions of daily living. 
We processed the smartphone video into an environment mesh and the IMU data and camera recordings into a human body pose estimate, body variables, and action labels. We collected movements with acted-out PD behaviour from two members of the research team (one developer and one clinician) and real movements from a person with PD.

\subsection{Data Processing}

\subsubsection{Environment Reconstruction}
\label{data_environment_reconstruction}

An environment reconstruction provides important context for understanding body motion and potential motor deficit triggers. PD-Insighter uses a mesh of the environment generated from the RGB camera data to show where the person was in 3D space. Given a video of the room, we use RealityCapture photogrammetry software \cite{noauthor_realitycapture_nodate} to create a dense point cloud and a texture map from video data to create a 3D mesh. We scale the environment using the real-world length of a physical referent in the space.

 \subsubsection{Action Recognition}
 \label{action_recognition}
 Action labels are start and end frames that indicate the kind of actions occurring at a given timepoint (e.g., walking). These labels give clinicians crucial context for analyzing body movement and isolating relevant timepoints of interest. Our test datasets use manual labels for higher accuracy and to restrict our actions to activities of relevance for our collaborators, but state-of-the-art methods in computer vision could also be applied to automate action labeling \cite{lin_learning_2022, wang_attentive_2019,demrozi_comprehensive_2023,kolkar_human_2023}.

During our design iterations, clinicians discovered that certain events of interest could be approximated using body variables. Specifically, clinicians found that freeze events could be approximately characterized by instances when the feet were below the pelvis for longer than a second during a walking event. We defined a mathematical filter to automatically label candidate activities matching this criteria. In future investigations, automated action labeling could incorporate such action approximations for condition-specific events with sparse training data, such as freezes.

\begin{figure}
  \centering
  \includegraphics[width=\columnwidth]{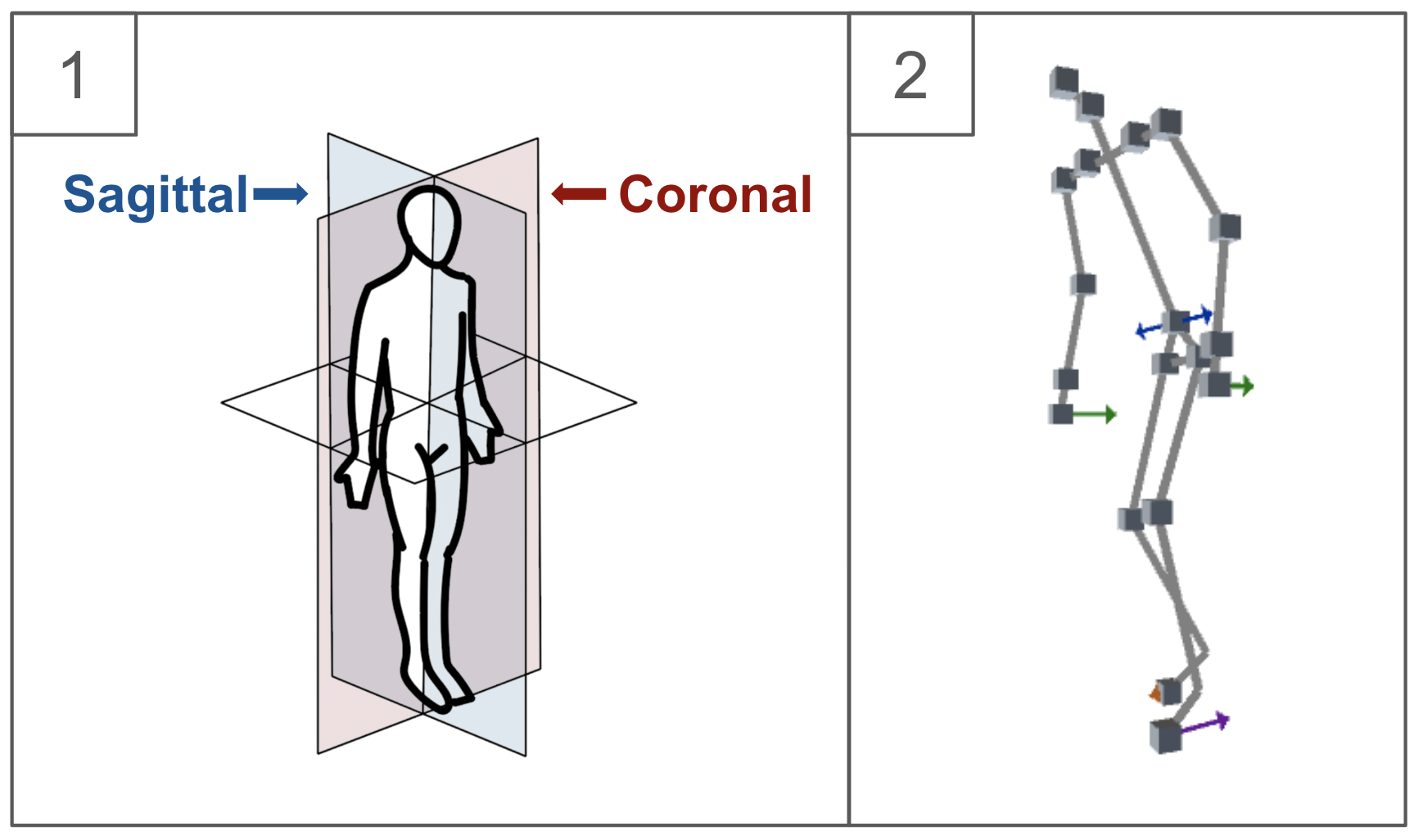}
  \caption{\textbf{Calculating Body Data}:
  1) Cardinal planes of motion, with the sagittal plane in blue and the coronal plane in red. 2) Skeleton rendered from human body pose coordinates, embedded with arrows encoding the body variables "arm use", "foot position", and "weight shift".
}
\Description{Figure 4.1 shows a person with 3 orthogonal planes intersecting at the pelvis of the person. One of those planes labeled "Saggital" indicates movement in the forward/backward direction . Another plane is labeled "Coronal" which indicates movement in the side to side direction. Figure 4.2 shows a grey skeleton with colored arrows stemming out of the skeleton's pelvis, hands and feet. }
    \label{fig:body_data}
\end{figure}

\subsubsection{Body Variables} 

\label{body_variables}

PD-Insighter requires body pose data to calculate and display joint displacements and angles reflecting shifts in motion and potential motor deficits (see \S \ref{sec:tasks}) as well as to reconstruct the person's body for immersive detailed investigation. We used EasyMocap \cite{noauthor_zju3dveasymocap_2023} to process IMU and camera video data in our collected datasets to compute pose coordinates for 22 joints per frame. While PD-Insighter can take an arbitrary number of body coordinates for reconstruction, it requires pelvis, hips, neck, hands, and feet coordinates to calculate the target body variables. 

PD-Insighter processes the body motion data to estimate four critical body pose metrics for detecting the deficits summarized in Section \ref{sec:deficit_tasks}: trunk angle, arm use, foot position, and weight shift. We compute relevant body variables using the positions of the pelvis ($P$), hip ($H$), neck ($N$), hands ($HA$), and feet ($F$). For a given joint position $J$, $j_i$ is the position of that joint at frame $i$ and $j_i^s$ is the side of the body the joint is on (left $l$ or right $r$).

Clinicians described these body variables based on the direction the limb or joint is moving in, called the \textit{cardinal planes of motion}. Movement in the \textit{sagittal plane} involves forward/backward motion, while the \textit{coronal plane} involves side-to-side motion (Figure \ref{fig:body_data}.1). We calculate the cardinal plane using a local coordinate system aligned with the pelvis $p_i$, assigning $\vec{y}$ to the vertical unit vector, $\vec{y} = \langle0,1,0\rangle$, $\vec{x} = p_i - h^l$, and $\vec z =\vec x \times \vec y $. We use this coordinate frame to compute the following body movement variables: 

\vspace{6pt}
\noindent\textbf{Trunk Angle:} The trunk angle $\theta \in \Theta$ is how far a person is leaning forward or backward in the sagittal plane at frame $i$. For all frames, we calculate a trunk vector $\vec{t} = p_i - n_i$ and find the angle between the trunk vector and $\vec{y}$ (i.e., a perfectly upright posture):  
$$\theta= \arccos (\frac { \vec{t_i} \cdot \vec{y}}{|\vec{t_i}|} )$$

\vspace{6pt}
\noindent\textbf{Arm Use:} The arm use $\rho^s_{\mathrm{arm}} \in  \Delta ^s_{\mathrm{arm}}$ is how far the hands are from the pelvis in the sagittal plane at frame $i$. For all frames, we calculate hand vectors $ \vec {ha^s} = p - ha$ and find the projection of the hand vectors in the sagittal direction $\vec{z}$:
$$ \rho^s_{\mathrm{arm}} = |\vec{ha_i^s} \cdot \vec{z}| $$

\vspace{6pt}
\noindent\textbf{Foot Position:} The foot position $\rho^s_{\mathrm{foot}} \in \Delta^s_{\mathrm{foot}}$ is how far the feet are from the pelvis in the sagittal plane at frame $i$. For all frames, we calculate feet vectors $ \vec{f_i^s} = p_i - f_i^s$ and find the projection of the feet vectors in the sagittal direction $\vec{z}$:
$$ \rho^s_{\mathrm{foot}} = \vec{f_i^s} \cdot \vec{z} $$

Unlike $\rho^s_{arm}$, clinicians want to know when the foot is in front or behind the pelvis, so we do not take the magnitude of the dot product to preserve the sign. The foot is in front of the pelvis if the dot product is positive and behind if negative. 

\vspace{6pt}
\textbf{Weight Shift:} The weight shift $\omega \in \Omega^s$ is a ratio dependent on how close the pelvis is to the feet coordinates at frame $i$. The closer the pelvis is to one side, the more weight that side is bearing. Using the feet vectors $ \vec{f_i^s} = p_i - f_i^s$, we find the projection of the feet vectors in the coronal direction $\vec{x}$: 
$$ \rho^s_{\mathrm{weightShift}} = |\vec{f_i^s} \cdot \vec{x}| $$

Given that clinicians focus these two vectors in relation to each other, we calculate ratios rather than displacement to represent weight shift:
$$ \omega^s = \frac {\rho^s_{\mathrm{weightShift}}}{\rho^l_{\mathrm{weightShift}} + \rho^r_{\mathrm{weightShift}}}$$

Ratios emphasize small but relevant imbalances. If both feet are close to the body but the weight is closer to the right, the small difference between the two sides would minimize the imbalance, whereas a ratio would highlight the differences. Ratios also reflect when there is balance or imbalance. For example, when $ \omega^l = 0.5$, the weight shift is mostly balanced, but when $\omega^l = 0.7$, the weight shift has a heavy lean to the left.

\section{Overview Dashboard}
\label{overview}

\begin{figure*}
  \includegraphics[width=\textwidth]{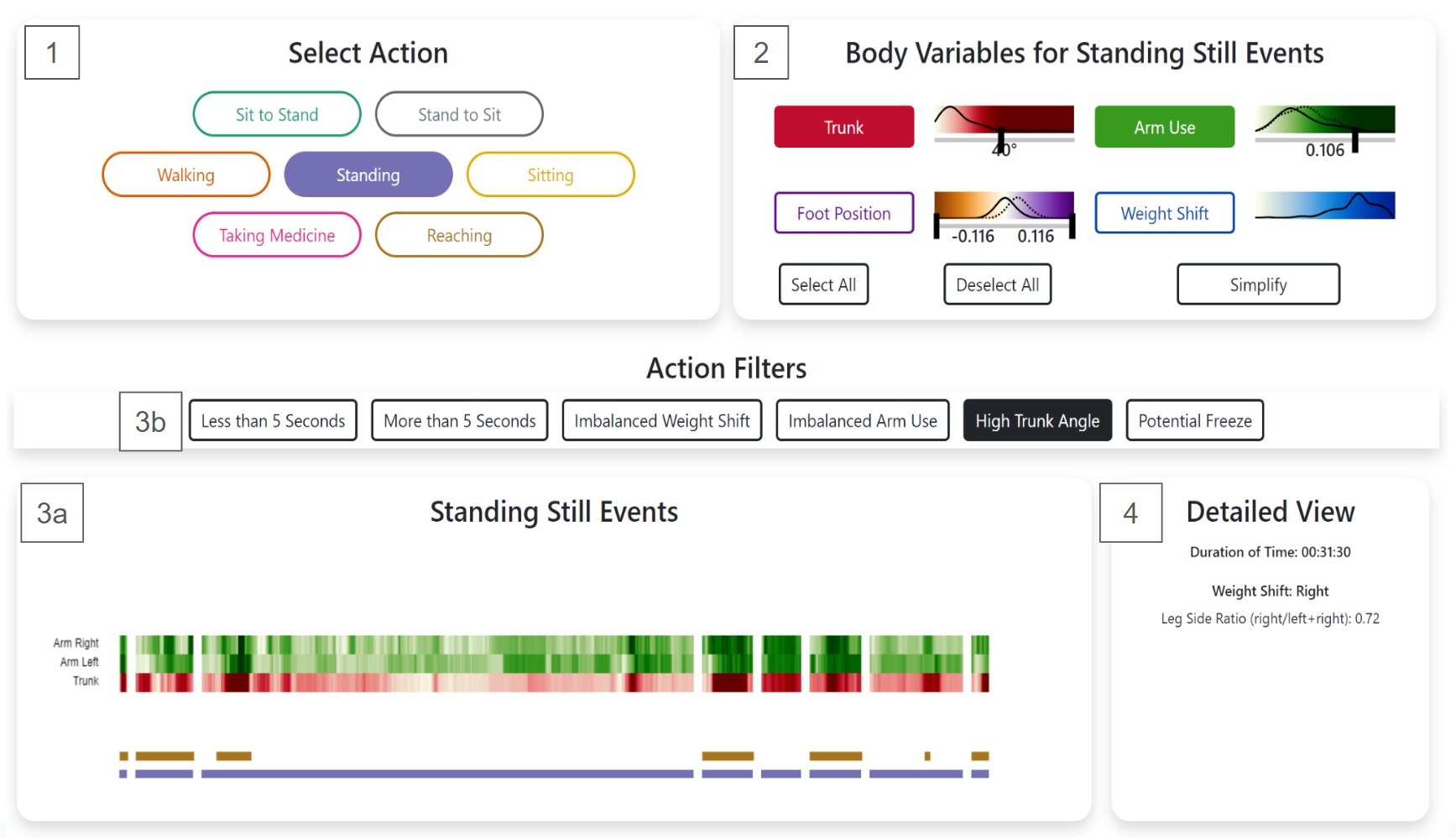}
  \caption{ An example use case of our \textbf{Overview Dashboard}: A clinician selects the Standing action (1) and Trunk and Arm Use body variables (2). In the Timeline window, PD-Insighter displays Standing  events and Trunk and Arm Use temporal heatmaps (3a), with additional filters available for further investigation, and the High Trunk Angle filter selected (3b). Statistical metrics summarize the displayed data (4).  
}
\label{fig:overview_dashboard}
\Description{Figure 5 shows the Overview Dashboard made from 5 sections. 1) "Select Action" contains 7 buttons labeled with different actions, with the "Standing" button selected. 2) "Body Variables for Standing Still Events", shows four buttons "Trunk", "Arm Use", "Foot Position" and "Weight Shift". Each of these buttons are next to a color ramp with a line graph distribution overlaid on top of the color legend. "Trunk" (colored red) and "Arm Use" (colored green) are selected. 3a) "Standing Still Events" shows a green temporal heatmap and a red temporal heatmap, corresponding to the selected trunk and arm body variables. Underneath are brown and light blue horizontal bars, corresponding to the Reaching and Standing Still actions respectively. 3b) "Action Filters" shows multiple filters, with "High Trunk Angle" selected. 4) "Detailed View" includes "Duration of Time" and "Weight Shift: Right", showing statistic metrics related to the selected data. }
\end{figure*}

PD-Insighter uses an overview-first, details-on-demand workflow. The Overview Dashboard supports early-phase analysis 
of critical actions and events 
(Figure \ref{fig:overview_dashboard}). Clinicians use Action Selection (\S \ref{action-filters}) and Body Variable Controls (\S \ref{body-sliders}) to drill down into data pertaining to relevant events in the Timeline (\S \ref{data-viewer}). They can see statistics of the selected data in the Detailed View (\S \ref{summary}). and replay these events using a video thumbnail or the Immersive Replay (see \S \ref{immersive}).

\subsection{Action Selection}
\label{action-filters}

To support identification and filtering by action (Task 1), 
selecting an action enables clinicians to focus their attention and analysis on particular actions of interest. Predefined buttons display time periods when the patient performed the selected action (Figure \ref{fig:overview_dashboard}.1). Our initial prototype presented all motion data over time with accompanying actions, using only temporal sliders for navigation. However, clinicians felt this presentation was overwhelming and preferred to use actions to isolate time segments of interest.
 
 PD-Insighter represents each action event with a horizontal bar. The bar's width is proportional to the duration of the event, preserving temporal ordering and showing which events are shorter or longer. We map each action to a distinct color in a categorical palette from Color Brewer \cite{harrower_colorbrewerorg_2003} to provide a compact action representation on the Timeline window. Each action type is presented in its own stacked row so clinicians can view simultaneous actions.
 For example, a high trunk angle while standing might be concerning to a clinician, but a high trunk angle while standing and reaching would not be concerning since the reach would explain the presence of the increased trunk lean (Figure \ref{fig:overview_dashboard}.3a). 

\begin{figure*}
  \includegraphics[width=\textwidth]{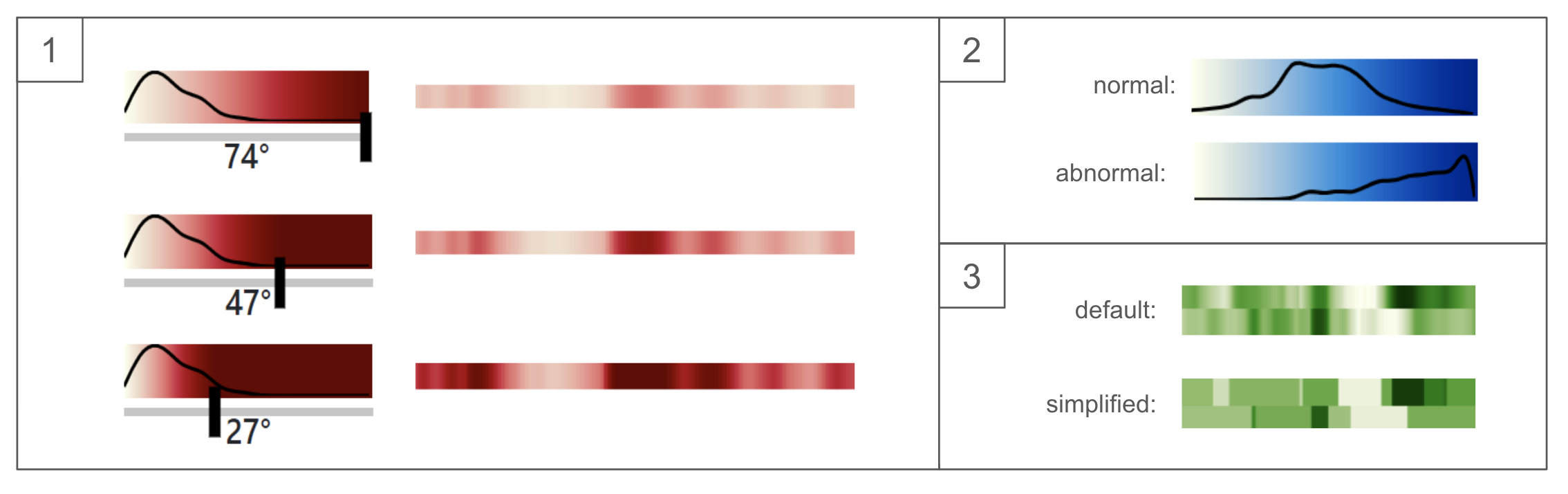}
  \caption{\textbf{Body Variable Representation:} Moving the body variable legend slider changes the temporal heatmap to emphasize different motion thresholds (1). The distributions overlaid on the sliders can help clinicians quickly identify 
  normal and abnormal distributions in weight shift balance and other body variables (2). The default temporal heatmap has more details and variations compared to the simplified heatmap that emphasizes outliers (3).} 
  \Description{Figure 6.1 shows a chi-square line graph distribution overlaid on a red color ramp.  A slider is under the graph and color ramp. Besides the slider is a red heatmap. As the slider moves to the left getting closer to the hump of the curve, the heatmap's values gets darker. Figure 6.2 shows two line graphs. The line graph labeled "normal" shows a gaussian curve. The second line graph labeled "abnormal" shows a curve heavily skewed to the right. Figure 6.3 shows two temporal heatmaps. The heatmap labeled "default" shows a lot of color with light and dark values. The heatmap labeled "simplified" is a simplified version of the "default" heatmap, showing much less variation with bins of white, green, and dark green values.
}
    \label{fig:color_sliders}
\end{figure*}

\subsection{Body Variable Controls}
\label{body-sliders}

To help clinicians discover motor deficits (Task 2), 
the Body Variable Controls allow clinicians to display relevant motion data through body variable buttons and sliders (Figure \ref{fig:overview_dashboard}.2).
Clinicians use body data to efficiently compare the action events and determine body motion patterns, potential deficits, outliers, and specific events for further study. We display our body variables---trunk angle, arm use, foot position, and weight shift---with four temporal color-encoded heatmaps to provide a compact representation of the data over time and to enable clinicians to quickly spot outliers. 
We assign each body variable a color ramp from Color Crafter \cite{smart_color_2019}
using distinct hues for each variable. 
Body variable color legend sliders allow the clinician to change the maximum value of a color ramp to make desired value ranges more salient. 

Overlaid on corresponding color legends, the line graph distributions for each variable show the frequency of the body variables across the dataset. These distributions allow clinicians to more rapidly tailor the color encoding in the Timeline to specific value ranges (Figure \ref{fig:color_sliders}.1).
Given the importance of left-right relationships in the feet and hands (see \S \ref{sec:tasks}), we pair their distributions, distinguishing the left from the right using a dotted line. Distributions can also summarize balance and posture for selected time segments to assist in global motion analysis. For example, when a patient has a balanced weight shift and upright posture, the distributions show a Gaussian curve and equivalent left and right measures. Unexpected distributions, such as an uncentered weight shift curve, high trunk angle values, or unmatching left and right distributions, can reflect poor balance, posture, or gait (Figure \ref{fig:color_sliders}.2).

\begin{figure*}
  \includegraphics[width=\textwidth]{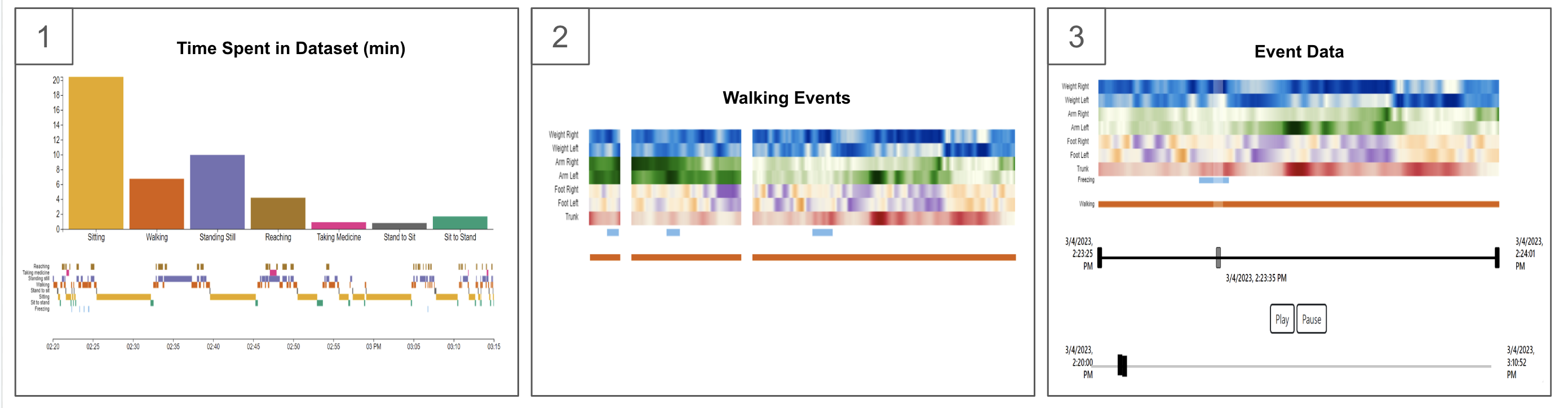}
  \caption{\textbf{Timeline:} When the clinician first opens the Overview Dashboard, they see the Action Summary and Action Timeline (1). Then, based on the action and filters selected, temporal heatmaps with encoded body variables for the selected action events (e.g., Walking events with potential freezes) (2). Once the clinician selects an action, the timeline zooms in to focus on the selected event, with range sliders for additional navigation (3). 
}

 \Description{Figure 7.1 displays a bar chart reflecting the amount of time the patient spent performing each action. Beneath the bar chart, a colored horizontal bar indicates the timeline of when actions were performed during the capture period. Figure 7.2 depicts 'Walking Events' with three colorful temporal heatmaps for three walking events. Light blue bars represent times when PD-Insighter predicts a freeze during a walking event. Figure 7.3 shows 'Event Data', featuring one of the heatmaps from Figure 7.2 enlarged, along with two range sliders and 'Play' and 'Pause' buttons.}
    \label{fig:data_view}
\end{figure*}

\subsection{Timeline}
\label{data-viewer}

The Timeline displays detailed body 
and action data over time based on the clinician's selections using a series of linear heatmaps (Figure \ref{fig:data_view}). 
When no actions are selected, the clinician sees 
a concise synopsis of the patient's behaviors and movements throughout the dataset (Figure \ref{fig:data_view}.1). The Action Summary uses a bar chart to summarize the time spent in each action 
and reflect the patient's general activity levels. For example, if the clinician sees the patient sitting most of the time, they may suggest that the patient walks around and stands more frequently. The Action Timeline uses a color-coded timeline to visualize the patient's actions chronologically, reflecting when and how long potential events of interest took place. This timeline helps clinicians understand sequences of related actions as well as general patterns of behavior, such as when the patient took their medicine, how long they were sitting, and if they were walking for longer or shorter periods of time.

Clinicians begin their analysis by selecting an action from the Select Action window or Action Summary chart. The Timeline window then updates to display a set of stacked temporal heatmaps reflecting the distribution of each body variable over time, with whitespace between discrete action events (Figure \ref{fig:data_view}.2). 
Additional temporal and body filters (e.g., More than 5 Seconds, Imbalanced Weight Shift) shown in Figure \ref{fig:overview_dashboard}.3b allow the clinician to filter even further and center their attention on events that fit their analysis goals. For example, the "Potential Freezes" filter was developed from mathematical calculations using the foot position, described in Section \ref{action_recognition}. Using actions and body variables to drive data filtering allows clinicians to focus on semantically meaningful sets of movements. 

The large number of frames in a dataset 
can introduce clutter and 
make it difficult 
for clinicians to find outliers and patterns in the data. PD-Insighter provides a toggle in the Body Variable Controls to simplify the display to emphasize large shifts in motion data. Based on prior work in large-scale data aggregation \cite{albers_sequence_2011, correll_visualizing_2011}, our simplification feature bins data into mean bins for values within one standard deviation from the mean, outlier bins for values more than one standard deviation from the mean, and averages the values in each bin (Figure \ref{fig:color_sliders}.3). By separating the data into these bins, we preserve and emphasize relevant outliers. 

Once the clinician finds an event of interest, they can click on that action event to focus exclusively on data for that event (Figure \ref{fig:data_view}.3). Two range sliders---a global slider with time relative to the entire capture and a local slider with time relative to the specified event---control the time segment in the temporal heatmap. The global slider can expand the 
Timeline view to show what happened before or after the selected time segment, as well as show where the selected time segment is with respect to the entire dataset. The local slider enables more precise navigation for zooming into the data. The slider also presents the start and end time to contextualize when in the day this happened. 
Once a clinician finds a moment they want to study further, the clinician can use the play/pause buttons or play slider to fetch video and reconstruction data to play in the Detailed View or Immersive Replay to explore an event in context (Task 3).

\begin{figure*}
  \includegraphics[width=\textwidth]{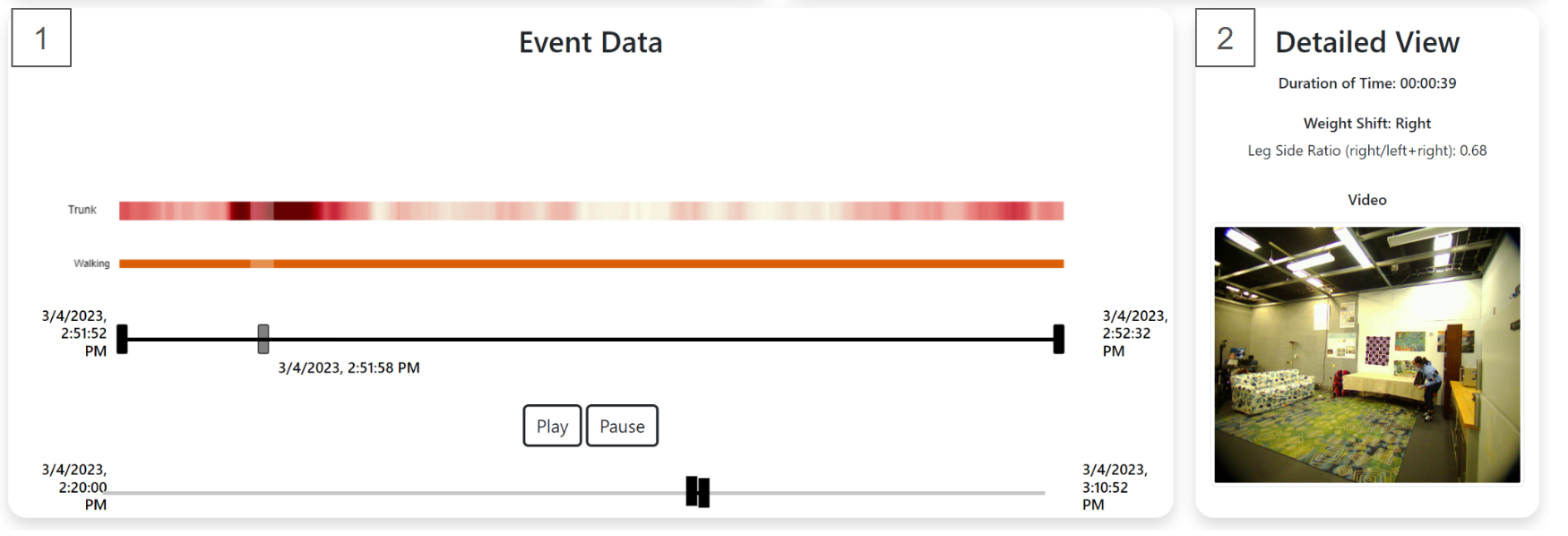}
  \caption{\textbf{Detailed View with Video:} When the clinician selects an event, the heatmap is shown in the Timeline window (1), and the Detailed View is updated (2). The Detailed View shows a video replay controlled by either the play/pause buttons or the play slider. 
}
    \Description{Figure 8 shows a subsection of the Dashboard, with two sections shown. 1) Underneath a red temporal heatmap labeled trunk, there is a horizontal orange bar labeled "Walking", two range sliders, and two buttons "Play" and "Pause". 2) "Detailed View" showing "Duration of Time", "Weight Shift: Right", and "Video" with an image of a person unbalanced. }
    \label{fig:video_display}
\end{figure*}

\subsection{Detailed View}
\label{summary}

The Detailed View presents statistical metrics and video replays for selected data. Based on clinician feedback, weight shift balance 
is a critical measure that 
can indicate when patients favor a particular side and potentially dangerous imbalances that can lead to a fall. The \textit{Weight Shift} metric provides an average weight shift side, weight shift measures, and a text-based description of the weight shift. While access to specific measures provides precise comparisons, clinicians indicated that the raw measures could be difficult to interpret. We added text descriptions (e.g., "strong right," "slight left," or "balanced") to support this analysis. 

Some metrics are more useful based on the selected time segment(s), so the Detailed View displays different metrics based on the current selection.
For example, when no action is selected, clinicians can look at the overall percentage of time sitting,
which is useful since large periods of inactivity can encourage motor decline. When the clinician selects an action, the \textit{Duration of Time} metric shows how much time was spent performing that action, which can shape 
interpretation. 
For example, two freezes in one hour is more concerning than two freezes across two days. 

When the clinician selects an action event to closely study, the Detailed View allows clinicians to replay 2D video data (Figure \ref{fig:video_display}.2). Video data from our data capture provides ground truth to compare with the temporal heatmap to confirm if outliers are from motor deficits or 
errors in data collection or processing. Further, the 2D video segment can provide some context for interpreting motion variables, supporting Task 3. However, clinicians felt that the video replays only offered limited perspective into body movement. 
To support more detailed replay, clinicians can launch the Immersive Replay to explore the target movements.

\begin{figure}
  \includegraphics[width=\columnwidth]{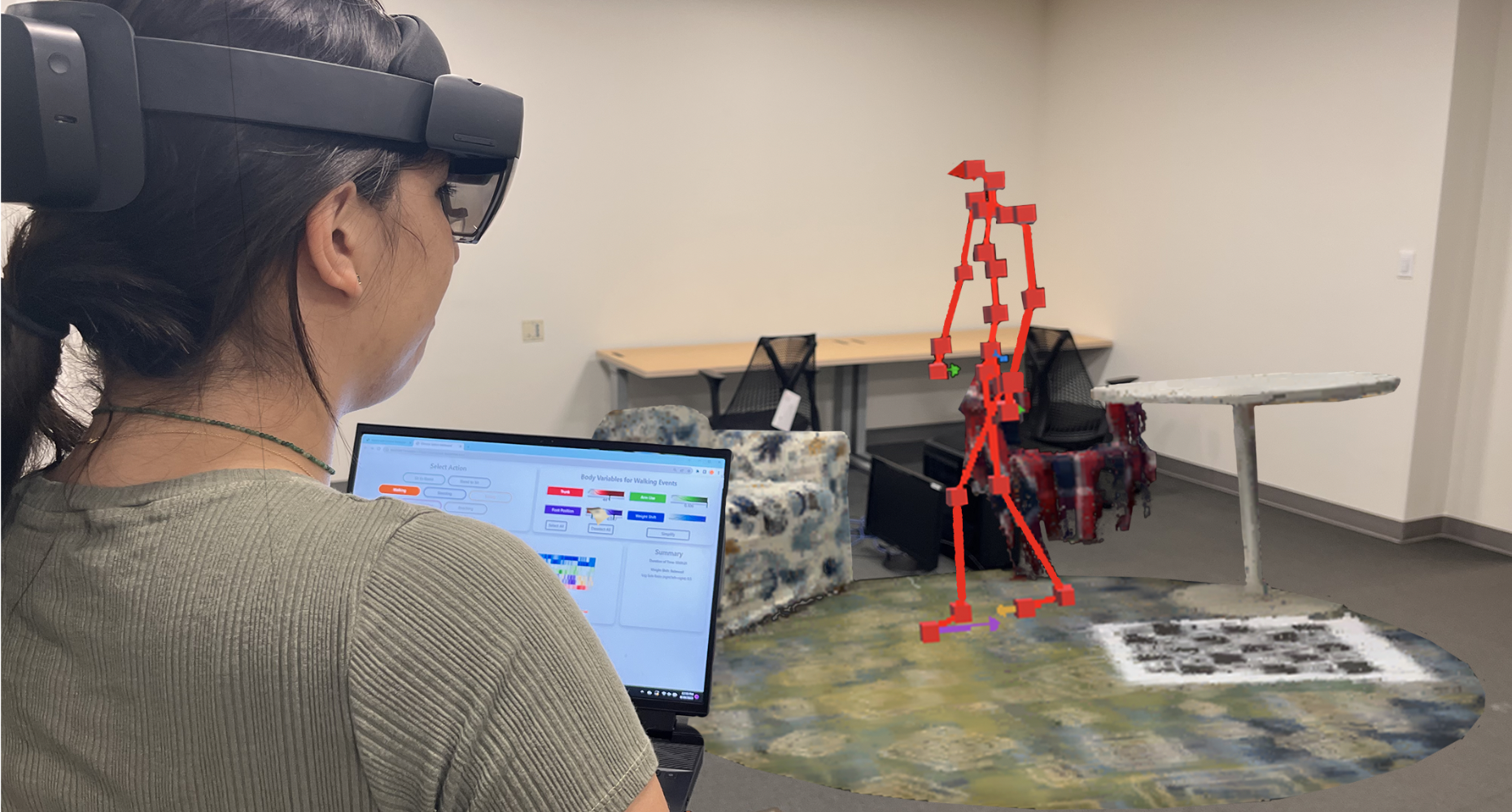}
  \caption{\textbf{Immersive Replay:} The clinician can view the Overview Dashboard on a desktop display and the Immersive Replay in AR. The Immersive Replay shows a body skeleton with arrow encoding the body variables situated in an environment reconstruction masked with a 1.5 meter radius. The rug, sofa, and pedestal table are in the AR environment; the tan table and black chair are part of the clinician's environment.
}
\Description{Figure 9 shows an individual wearing an augmented reality headset. This person observes a digitally-rendered skeleton within a virtual environment, simultaneously managing the Overview Dashboard on a laptop.}
    \label{fig:immersive_replay}
\end{figure}

\section{Immersive Replay}
\label{immersive}

To contextualize motor deficits (Task 3), 
clinicians can interact with the Immersive Replay to study a reconstruction of a time segment of interest in an AR headset. Our Immersive Replay consists of a skeleton rendered using the body pose coordinates situated in a local environment reconstruction.

We use AR to display this reconstruction as the analytical workflows of AR mirror critical practices that clinicians use with patients during in-person therapies. Clinicians are accustomed to physically maneuvering and examining a patient's body in a clinical setting. AR allows clinicians to view a life-size reconstruction of the patient within the context of the environment. The clinician can then walk around the reconstruction to view the motion from multiple angles and deeply attend to movement biomechanics which directly inform therapies. Further, the enhanced depth cuing provided by AR rendering enhances clinicians' abilities to reason about body position compared to viewing desktop-based reconstructions \cite{heinrich2021estimating}. 
 
Clinicians move back and forth between the dashboard and reconstruction without having to remove the headset (see \S \ref{headset}). Our reconstruction pipeline can adapt to various privacy needs by using wireframe skeletons and allowing people to turn off the environmental context when necessary. 

\subsection{Body Reconstruction}
\label{body_reconstruction}
The patient's body is reconstructed using a simple skeleton. We use this design rather than a more photorealistic reconstruction because it is visually simpler, provides additional patient privacy, decreases occlusions, enables situated visualizations of relevant body metrics (e.g., motion vectors), and requires less memory. We colored the skeleton bright red to distinguish the body from most naturally-occurring environments. 

We visualize body variables using arrows at relevant coordinates to show the associated displacements in physical space. Arrow magnitude and direction correspond to the magnitude and direction of the body variable at that frame. The colors of the arrows match the colors of the body variables in the Overview Dashboard to connect the Immersive Replay with the studied body variables. 

\subsection{Environment Reconstruction}
\label{environment_reconstruction}
We situate the skeleton within an immersive reconstruction of the environment extracted from video data (see \S \ref{data_environment_reconstruction}). PD-Insighter centers the reconstruction on the skeleton, showing a 1.5 meter radius window of the environment around the body. 
Original iterations of the Immersive Replay involved the body reconstruction moving in a stationary environment. However, clinicians noted that any analysis would likely happen in their office before a patient visit. The clutter and limited space of the clinician's office inhibit walking around large reconstructed environments, whereas centering on a smaller local window means clinicians can more easily walk around the local reconstructed environment to study contextualized motion patterns in detail. 

\subsection{Head-Mounted Display}
\label{headset}
PD-Insighter uses the HoloLens 2 headset to enable 
clinicians to seamlessly transition
between the Overview Dashboard and Immersive Replay by 
shifting their gaze between the desktop display and the immersive reconstruction. For easier mobility, clinicians can view the Dashboard on a laptop and physically move around the immersive reconstruction to look at 
the patient's movements from
multiple views. Early iterations of the system used Oculus Quest 2 to display both the dashboard and reconstruction, but clinicians found using the device controls for interaction cumbersome. We switched to see-through AR to allow clinicians to use the desktop to interact with the Overview Dashboard and provide more comfortable physical interaction with the Immersive Replay. 

\section{Evaluation}

We evaluated PD-Insighter through a think-aloud study with six rehabilitation specialists. Five participants had direct experience with PD, while the sixth worked with people with stroke facing similar mobility challenges. Two therapists had also served as expert collaborators during early design iterations, whereas four had not seen the system before. For transparency, we label data from early-phase collaborators with C1 and C2, and other participants with P3, P4, P5, and P6.
Drawing on our task analysis, we focused our evaluation on PD-Insighter's abilities to support  1) overall understanding of a patient's motor capabilities, 2) finding specific moments of interest that correspond to potential motor deficits, and 3) studying these moments of interest 
with context. 

The therapists used PD-Insighter to analyze a 50-minute dataset consisting of four capture segments containing  the seven key actions from our task analysis. During data collection, we intentionally captured a diverse range of relevant deficits through the research team
and patients with PD performing these actions. We envision a future with potentially days or weeks worth of data, but current technologies limit that capability at present. We use a 50-minute composited capture as it reflects likely data composition in practice. Current captures are often done in-lab over multiple visits. Patients at home will likely only record data in segments to avoid fatigue from wearable devices, to recharge the battery of devices, and to protect privacy during sensitive actions like using the restroom. While PD-Insighter can support analysis with longer captures, participants noted that 
50 minutes
of data was long enough to understand general movement patterns and include a range of key specific movements, such as falls and freezes, for evaluation. 

First, we introduced the system by describing the tool's basic functionality and body variables. We then asked participants to put on the HoloLens 2, analyze data to identify potential motor deficits that might influence treatment, and make general observations about the patient's movement patterns. We asked participants to describe their observations, intentions, and decisions as they worked with the tool. 
We gathered additional feedback in an exit interview. Each session approximately lasted one hour. 

\subsection{Overall Movement Patterns}

All participants engaged with the opening Action Summary and Timeline (Figure \ref{fig:data_view}.1), which C1 described as ``necessary and grounding,'' and would frequently return to after studying an event. C2 and P3 used the Action Summary bar chart and the Percentage of Time Sitting metric to see if the patient was active or inactive. 
P6 thought the Weight Shift statistic would be helpful because “if you don’t have time, you can really quickly see a snapshot that tells how [the patient] has been doing."
P3 remarked that the feature would be helpful for clinical analysis since ``the amount of sitting tells a lot about how active  someone was, which is helpful even for stroke patients, not just Parkinson's.''

Participants also focused on using the body variable distributions to identify patterns in motion across actions. C1 and C2 were interested in whether the patient used both arms or just one arm while reaching because ``Typically in Parkinson's, there can be an asymmetric representation [between arms] because they might be compensating. [For example,] if there is a tremor, they could be hesitating to use that hand'' (C1). By studying the Arm Use heatmaps for reaching events, C1 and C2 both found that there were several reaching events with single dominant hand and bi-manual reaching, indicating ``no concern'' (C1; Figure 10.1).

The Action Timeline was ``useful for thinking of the context of what [the patient] was doing through the day. When are they taking medicine? Are they walking for shorter or longer?'' (C1). P4 also remarked
``It is helpful to know at what point they are taking medicine because it has a huge impact on their motor function. It takes about an hour for medicine to kick in.''

\subsection{Finding Moments of Interest}

\begin{figure*}
  \includegraphics[width=\textwidth]{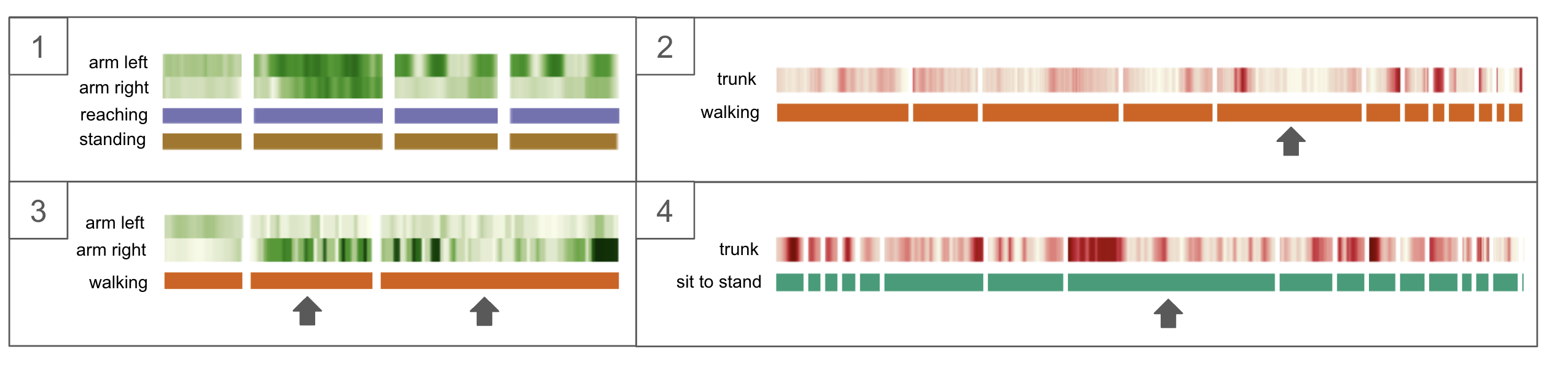}

  \caption{\textbf{Evaluation--Finding Moments of Interest:} Clinicians found both bi-manual and single-hand reaching, indicating that the patient is comfortable reaching with both hands (1). Dark red bands indicating a high trunk angle increase drew the participant's attention, leading to a discovered fall event (2). Participants found events with imbalanced arm swing while walking (3). Long events with alternating trunk can indicate a struggle to stand up from a chair (4).
}

\Description{Figure 10.1 displays green heatmaps labeled 'arm left' and 'arm right' during reaching events. Some heatmaps show dark green for both the left and right arms, while others show dark green bands only for the left arm. Figure 10.2 depicts red heatmaps labeled 'trunk' during walking events, with an arrow pointing to one of the heatmaps that features a dark red area. Figure 10.3 presents green heatmaps labeled 'arm left' and 'arm right' during walking events. Arrows point to two heatmaps, where the right arm's heatmap is nearly white, and the left arm's heatmap shows dark green bands. Figure 10.4 illustrates red heatmaps labeled 'trunk' during sit-to-stand events, with an arrow pointing to the longest heatmap that contains alternating dark red and light red bands.  }
    \label{fig:finding_moments}

\end{figure*}

 All participants used the Select Action feature when 
 initiating a new line of inquiry. Once participants picked an action, they chose body variables of interest to display. Participants frequently remarked how various actions impacted their expectations for the body variables. For example, ``high trunk is concerning for walking, but not for reaching'' (C2), and ``Arm use while walking will be different than arm use while reaching'' (C1). 

Participants commonly focused on events with dark-colored outliers to find specific instances of motor deficits. For example, C1, C2, P3, and P6 were interested in moments with imbalanced arm swing while walking since ``swinging with the left arm and not the right arm is characteristic of Parkinson's'' (C1). When they selected Walking, Arm Use, and the Imbalanced Arm Use filters, 
participants were interested in instances where the right arm heatmap was close to white, and the left arm heatmap had dark green bands (Figure 10.3), reflecting extended arm swing. Participants felt confident that imbalanced swings were occurring without viewing a replay. 

P3, P4, and P5 clicked on a Walking event with dark red bands indicating a high trunk angle and found a freeze (Figure 10.2) missed by the Potential Freeze filter. 
C2 clicked on a Standing event with a dark red band and found a fall. 
P6 clicked on a Sit-to-Stand event with dark a red band because “a lot of falls are happening from leaning too far forward.” C1 clicked on a Sit-to-Stand event with multiple dark red bands, which reflected an event where the patient was trying to lift their weight out of the chair and struggling (Figure 10.4). P3 and P4 used the blue heatmaps to determine weight shift imbalances. P4 remarked that ``in general, the weight shift is more on the right than the left,'' and was interested in this imbalance during walking events because it showed the patient ``having a prolonged period of single limb stance.'' 

P5's navigation was heavily driven by color contrast to find potential deficits, ``Really what I am looking for, just for the sake of it, is a moment where one (heatmap) is darker than the other (heatmap).'' When looking at Stand-to-Sit events, P5 had all body variables selected, saying, ``I am going to click all of them until I can decide what I want to look for,'' and chose the event with the darkest trunk angle. Then, P5 looked at the Foot Position and chose the event with the darkest purple. This behavior indicates PD-Insighter's ability to support data analysis based on unexpected and significant shifts in different parts of the body relative to others, highlighting outliers and potential deficits.  

Participants also used time duration to find moments of interest. For example, C2 wanted to see if the patient abruptly fell in their seat, indicating a ``failure to react to a change in position.'' C2 selected the Stand-to-Sit event with the shortest width and found the patient fell into their seat. P3 and P4 were interested 
in long Sit-to-Stand events indicated struggles standing up. P3 clicked on the More Than 5 Seconds filter to find longer events, whereas P4 looked at the width of the bars to ``see if they are getting better (at sitting down) over time.'' Both participants clicked the same long Sit-to-Stand event that showed a struggle to stand up (Figure 10.4). 

Filters helped decrease the number of events shown
and focus participants' explorations. 
All therapists clicked on the Walking and Potential Freezes filters to try to find freezing gait patterns. PD-Insighter showed three different walking events, with light blue bars indicating potential freeze moments (Figure \ref{fig:data_view}.2). By using the immersive replay to watch the reconstruction and see if there were body leans showing the patient trying to 
start walking but struggling, it took clinicians less than one minute to find that one of these events was a proper freeze.

\subsection{Contextualizing Moments of Interest}

\begin{figure*}
  \includegraphics[width=\textwidth]{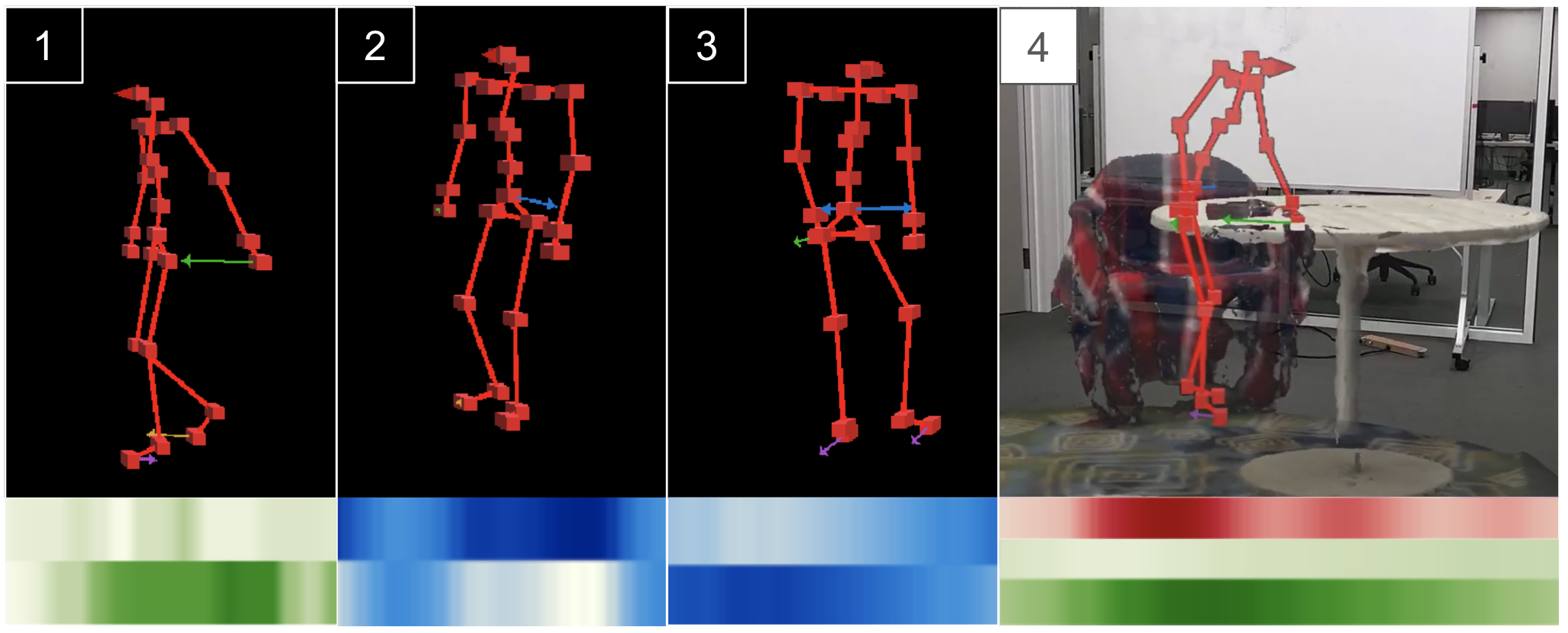}
  \caption{Evaluation--Studying Moments of Interest: Clinicians found imbalanced arm swing while walking (1), weight shift imbalance during a freeze event (2) and while turning (3), and instances of `furniture walking' (4). 
}
\Description{Figure 11.1 shows a red skeleton with its arm extended backwards, and a green arrow extending from the hand, pointing forward. Figure 11.2 shows a red skeleton with a blue arrow extending from the pelvis, pointing to the left. Figure 11.3 shows a red skeleton with two blue arrows extending from the pelvis; the one pointing to the left is much shorter than the one pointing to the right. Figure 11.4 shows a red skeleton with its hand on a virtual table. Beneath all these figures are colored heatmaps corresponding to the magnitude of the arrows over time.}
    \label{fig:evaluation-immersive}
\end{figure*}

While participants had access to the ground-truth video feed in the Detailed View, they primarily used the Immersive Replay for watching body motion because they felt ``the video is too constrained'' (P3) due to the inability to move around the body and look around any occlusions. The replay in AR was helpful because “angles matter, I need to orient myself to see [body parts of interest]” (C1). All participants found the headsets easy to use with “no discomfort” (P6). 

Therapists used the Immersive Replay to confirm or deny the existence of a motor deficit in an event of interest. 
For example, when studying standing events, C2 saw a dark red band while reaching and ``suspected it was just a normal reach since when reaching a high trunk angle is common.'' However, the red was significantly darker than the other trunk leans while reaching, leading C2 to further investigate. When watching the reconstruction, they found the reach led to a fall. The reconstruction ``helped me see how hunched they were, that it was a fall and not just a reach.'' While looking for freezing events, all participants used the reconstruction to determine if a freeze occurred. When they saw the skeleton leaning back and forth and shuffling before stopping, they accurately characterized the movement as a freeze. Once discovering the freeze, C1, C2, and P4 returned to the Overview Dashboard to study the event in more detail. For example, C1 wanted ``to see if the trunk increased, which is indicative of more risk for falls.'' In the Overview Dashboard, they looked at the trunk angle heatmap and found that the trunk ``looked higher but not substantially'' with respect to the rest of the data. 

Participants felt that the arrows encoding body variables in the Immersive Replay provided helpful clarity. When watching an event with an imbalanced arm swing, C2 said, ``The arrows give help with the direction. I can see the arm swing is more backwards than forwards.'' P4 said that ``the arrows make it clear that the arms are obviously not synchronized'' 
which leads to checking, ``Are they using their assistive device or furniture?'' (Figure 11.1). 

Therapists also used the blue weight shift arrows for further analysis. 
For instance, ``When [the patient is] turning while walking, I can see how the arrow changes to emphasize the weight bearing, and I can see more on one particular side.'' (P3) (Figure 11.3). P4 liked this visibility because ``turning is a time when people will fall. It's a good thing [for the patient] to shift their weight while turning.'' P4 used the weight shift arrows to study a freeze event to realize that the patient was ``trying to move one side but can’t initiate'' (Figure 11.2). After finding this imbalance in the arrows, P4 went back to the heatmap and saw that during the freeze event, the right side was dark blue, and the left side was white, confirming that the patient was placing their weight on their right side to try to move forward. 

P5 used the weight shift arrows to study a Sit-to-Stand event, saying, ``There's some bias on the right side, so [the patient] will do something like this when they are trying to get up,'' and acted out a swivel to demonstrate the incorrect way to stand up. P5 considered the advice they would give the patient 
``Are they building strength on that [less affected] side? Maybe we need to adjust that chair, add pillows, wedge a book under the rocking chair.''

The environment provided participants with important context for analysis. When studying a Walking event, C1, C2, P3, P4, and P6 were curious about the high trunk leans in the heatmap. When watching the reconstruction in the Immersive Replay, C2 said, ``This shows me that they are furniture walking [walking while using furniture as an assistive aid]. Before, in the heatmap, it looked like they were walking weird. I would have thought they were about to fall'' (Figure 11.4). P4 remarked, ``Yeah, that’s definitely helpful to have the environment there. Otherwise, it would be like, what are the arms doing so far apart here?'' when indicating a point of divergence between hand positions on the heatmap. C2 mentioned the importance of seeing the environment and furniture walking, ``When a patient knows their center of balance will be disturbed, I want to see interaction with the furniture to help stabilize.''

\section{Discussion}
 PD-Insighter is a hybrid desktop-AR visual analytics system for physical therapists to analyze body motion data for patients with Parkinson's Disease during activities of daily living. The system works towards a vision of processing long-term at-home body motion data to support better decision-making and diagnosis. Iterative interviews with clinicians elicited key tasks that shaped our system design. Here, we summarize preliminary outcomes from the implementation and evaluation of PD-Insighter to inform future work for studying body motion data in clinical therapy applications. 

\subsection{Data Representation}
Most ADL visualizations represent data through activity level and location, which does not give the context necessary for identifying and studying PD motor deficits. We discovered that combining action and body variables can indicate motor deficits and body pattern changes. Clinicians noted that the action and body variables were necessary for a holistic picture of patient movement.

Based on the range of patient needs, we developed action labels and body variables from existing clinical practices to assess patients. P6 appreciated the chosen action labels, "These features are known from clinical observation research to be associated for risk of fall." Different clinicians can use the tool based on their expertise or the needs of their patients. P4 stated, ``I like having independence to look at what I want to look at based on what is important to me,” which can be “influenced by patient’s needs and what they are reporting.'' P6 described how actions of interest can also be influenced by personal specialty, “I am clinically more interested in walking, sit-to-stand, and stand-to-sit. How someone is sitting or standing is less relevant to me.”

Our evaluation showed that action labels and body variables successfully indicated motor deficits such as freeze events, imbalanced arm swing while walking, struggle standing up, and struggle sitting down. Clinicians successfully understood body motion in terms of these new body variables and could identify when the data was reflecting improper motion. While various data representations exist for body motion, both action and body data can work together to enable 
scalability across time and provide the necessary context for deeper analysis. 

\subsection{Detecting \& Emphasizing Outliers}
Most moments of clinical interest for PD body movement last only a few seconds. Finding these moments in an hour or more of data is time cost-prohibitive with direct replay: clinicians only have a few minutes to prepare for a patient visit. Emphasizing outliers in the data using dark, saturated colors helped quickly draw clinicians' attention to potential deficits. 

At the beginning of the evaluation, P3 thought that the color encodings would be challenging for PTs unfamiliar with this representation. However, within a few minutes, they found that ``Using [the colors] gets intuitive very quickly'' and had no issue interpreting the data. P3 described the heatmap as ``very telling. I don’t need additional information to see the weight shift balance, a quick glance tells which side bears more weight.'' Participants found that the simplification feature made the heatmaps easier to use (P3, P4, P6). C2 found the body variable sliders helpful to ``adjust the contrast [so that] it's easier to see'' and the distributions helpful to ``guide where to move the sliders.'' 

Task-aware aggregation methods, such as the simplify feature, can highlight these outliers without removing important information, which is helpful for large and noisy data. While several approaches for aggregation exist, we found that clinicians needed methods that 
both preserves and emphasizes outliers to ensure that important moments of motor deficit are not lost.

\subsection{Fault Tolerance \& Filtering}
\label{fault_tolerance}
When we developed PD-Insighter, we used datasets with manual labels. 
To date no tools exist for automatically and reliably labeling the specific activities of interest 
and we believed that accurate action labeling was critical. While accuracy is desirable, we also discovered that PD-Insighter created a certain degree of fault tolerance in motion analyses. Actions or relationships between body variables support quick, coarse guidance toward events of interest, and verifying events is sufficiently quick 
that clinicians could readily confirm or dismiss events. For example, although our Freeze Event thresholds introduced several false positives, clinicians did not mind the inaccuracies because they could confirm the event in a few seconds by simply watching a replay. 

Salient misalignments in body variables helped clinicians capture significant instances that action labels may have missed. For example, participants quickly began looking for time spans in the heatmaps where body variables were high in one value and low in others, helping address potential false negatives in labeling. As clinicians worked with the tool, they generated heuristics, such as those for the Freeze Event filters, that gave coarse approximations of critical potential deficits and informed clinical thinking about the computational relationships between joints in different movements. This introspection may enable future metric development and on-the-fly filtering for working with biomechanical data. 

The combined fault tolerance and developed mathematical intuitions about deficits exhibited in the evaluation indicate that approaches like PD-Insighter may support mixed-initiative action classification. Specifically, if an automated classifier embedded in PD-Insighter provides preliminary action labels, clinicians may confirm or deny events through their analysis. 
This confirmation 
can provide clinicians with valuable patient motion insights and classifiers with additional training data. 
Improving overall classification feedback and enabling customization for individual movement patterns has important potential for future research. Exploring the potential for such a system is important future work. 

\subsection{Hybrid Interfaces for Motion Analysis}

While the abstract heatmaps helped clinicians discover that an unexpected motor pattern occurred, they typically cannot explain \textit{why} the observed body motion change happened. 
Replays helped clinicians distinguish between motor deficits, inaccurate measures or heuristics, and potential triggers. All participants felt the video replay was inadequate for this analysis and preferred the up-close immersive reconstruction. 
Five of the participants walked around the model at least once during the analysis to look at the body from multiple angles. C2 emphasized the importance of having access to various perspectives by pointing out ``It's hard to see someone leaning forward from a front view; I would want a side view, and here I can.'' Clinicians study a patient's body motion in person, and AR recreated familiar experiences for retrospective analysis and delivered a better sense of the scale of the movements. 

The body and environment reconstruction work together to provide the necessary context. AR provides crucial depth and scale to analyze body motion properly. Further, AR provides necessary contextual cues from the environment. Reflecting on a prior iteration of the system without the reconstruction, our collaborator C1 noted the replay was ``much more helpful with the environment. I can see what they are doing and why they are doing it.'' Clinicians used the replay to analyze object placement in reaching, walking with furniture assistance, and chair placement when sitting. 

The functional gains from combining both interfaces outweighed the trade-offs associated with context switching between platforms. While clinicians found video replays in the dashboard inadequate, early iterations of the system experimented with situating the Overview Dashboard in a full VR reconstruction. However, clinicians had difficulty interacting with the data. With AR, no participants noted difficulties transitioning between the Overview and Replay. 
C1 described the interface as “fairly seamless”; the reconstruction is “right there.” Despite having no prior experience with head-mounted displays, P6 described the AR headset as "good, awesome, better than I expected," because "it does not prevent you from seeing other areas in the real world."

With participants finding the headsets easy to wear and interact with, as well as useful for viewing the replay in multiple perspectives, our findings align with recent research on how immersive renderings can improve 3D data analysis \cite{lin_towards_2021}, emphasizing the importance of comprehending trade-offs in visualization displays for specific design scenarios \cite{ens_grand_2021}.

\section{Limitations \& Future Work}

While PD-Insighter's current design works for a few hours of capture, such as those stitched together from a series of in-lab observations, we envision a near future where non-intrusive wearables are capable of capturing days worth of data for clinical analysis between visits \cite{zhang_reconstruction_2023, cha_mobile_2021}. Such scenarios will provide a more comprehensive listing of movements to be analyzed, creating new scalability challenges for working with motion data. We anticipate that our aggregation and filtering paradigms will support this analysis, especially as real-world captures are likely to be significantly less event-dense on average; however, future work should explore novel paradigms for task-aware aggregation, including those informed by mixed-initiative approaches as discussed in Section \ref{fault_tolerance}. 

In our current system, we do not directly address privacy preservation for the patient beyond wireframing and environment toggles in the reconstruction. As 
the tool is designed for clinicians working directly with a patient, we assume the patient has given their permission for clinicians to collect and analyze their data.
However, real-world deployments 
will introduce privacy concerns. While approaches like lateral-effect photodiodes (LEPDs) \cite{yang_implementation_2002} can provide more privacy-aware capture, future research, especially in patient-driven methods, is necessary to better understand privacy needs, constraints, and agency. Methods to address privacy may include using privacy-aware approaches in data preprocessing, capturing data in only certain spaces for limited amounts of time, and the ability to blur faces in video captures.

More data will require automatic methods for labeling actions. Future iterations of action-driven analysis tools like PD-Insighter can use computer vision models for automatically labeling video data based on the action performed \cite{lin_learning_2022, wang_attentive_2019, demrozi_comprehensive_2023, kolkar_human_2023}. Future tools could also integrate a more comprehensive array of body variables or even allow clinicians to define body variables during analysis. For example, step length and number of steps per second are valuable metrics for understanding gait (C1, C2, P3, P6). While PD-Insighter's design can readily be expanded to work with more variables, future work should develop measures for understanding and computing these variables to identify those of the greatest clinical relevance. 

 PD-Insighter currently focuses on supporting data exploration by physical therapists. Ultimately, the physical therapist will want to communicate their findings, discuss challenging events, and provide suggestions to the patient. Therapists can experience difficulty relaying this feedback verbally to the patient.  Hybrid spaces may help with PD rehabilitation by mediating communication with the patient. For example, patients can walk through their movement patterns in AR annotated with feedback from the clinician. Patient-centered interfaces may also help patients play a more active role in their treatment and retain greater agency over their data. We plan to address potential challenges in older patients' data accessibility by consistently reviewing and adapting our design in future collaborations with patients. While such AR systems may be difficult to adapt in the present due to usability limitations, the ongoing advancement of AR hardware over time will enhance their feasibility.

These methods in PD-Insighter can shape tools in other domains interested in body motion data. Study participants noted direct applications of PD-Insighter for other motor conditions, such as stroke, or even general movement analysis, as in sports and motion training. While current movement visualization tools focus either on aggregate movements \cite{robben_identifying_nodate, mulvenna_visualization_2011, ploderer_how_2016} or brief, focused snippets of motion \cite{rashidi_mining_2010, boers_smart_2009}, we anticipate that these techniques may serve as a foundation for other movement-based analysis tools.

\section{Conclusion}

Clinicians currently have limited access to patients' movement data outside the clinic, despite its significant potential for improving the treatment of motor deficiencies like those seen in Parkinson's Disease. We designed PD-Insighter to enable insights into patients' daily movements to inform clinical practices associated with physical therapy. 
In iterative consultations with physical therapists, we found that action and body variable data representation, emphasizing outliers, 
hybrid interfaces, and immersive replays with situated encodings enabled clinicians to find and analyze relevant motor deficits in body motion data. Our approach works towards a future where clinicians can remotely monitor patients 
to optimize personal treatments, dramatically improving clinician decision-making and slowing the functional decline of PD and other medical conditions. 

\begin{acks}
The authors would like to thank Jim Mahaney for help with setting up a capture space with various pieces of furniture to simulate a home environment. This work was supported by the National Institutes of Health Award 1R01HD111074-01 and NSF IIS-2320920.
\end{acks}

\balance
\bibliographystyle{ACM-Reference-Format}

\bibliography{template1}

\end{document}